\documentclass{llncs}
\usepackage[utf8]{inputenc}
\usepackage[english]{babel}
\usepackage{shortcuts}
\usepackage{hyperref}
\usepackage{mathpartir}
\usepackage{stmaryrd}
\usepackage[matrix,arrow]{xy}

\newcommand{\emptypos}{\dag}
\newcommand{\lrule}[1]{[#1]}
\newcommand{\llplusl}{\llplus_L}
\newcommand{\llplusr}{\llplus_R}
\newcommand{\llwithl}{\llwith_L}
\newcommand{\llwithr}{\llwith_R}
\newcommand{\fpos}[1]{\mathrm{pos}(#1)}
\renewcommand{\wrt}{wrt~}
\newcommand{\initpos}[1]{\ast_{#1}}
\renewcommand{\vxym}[1]{\vcenter{\xymatrix@C=1.8ex@R=1.8ex{#1}}}
\renewcommand{\tile}[1]{\mathop{\diamond_{#1}}}

\newcommand{\async}{\mathop{\Uparrow}}
\newcommand{\sync}{\mathop{\Downarrow}}
\newcommand{\focleq}{\prec}
\newcommand{\deseq}[1]{\tilde{#1}}
\renewcommand{\leq}{\leqslant}
\newcommand{\fix}[1]{\mathop{\mathrm{fix}}(#1)}
\renewcommand{\tq}{\ |\ }
\newcommand{\leqP}{\leq_P}
\newcommand{\clop}[1]{{#1}^\bullet}
\newcommand{\hpos}[1]{#1^\circ}
\newcommand{\astrat}[1]{#1^\lightning}

\title{Focusing in Asynchronous Games}

\author{Samuel Mimram\thanks{CEA LIST, Laboratory for the Modelling and Analysis
    of Interacting Systems, Point Courrier 94, 91191 Gif-sur-Yvette,
    France. E-mail:
    \href{mailto:samuel.mimram@cea.fr}{\email{samuel.mimram@cea.fr}}. This work
    has been supported by the CHOCO
    (ANR-07-BLAN-0324)
    ANR project.
  }}

\institute{CEA LIST / École Polytechnique}

\date{}

\hypersetup{
  pdftitle={\csname @title\endcsname},
  pdfauthor={Samuel Mimram},
  unicode=true,
  colorlinks=true,
  linkcolor=black,
  citecolor=black,
  urlcolor=black
}

\begin{document}
\maketitle

\begin{abstract}
  Game semantics provides an interactive point of view on proofs, which enables
  one to describe precisely their dynamical behavior during cut elimination, by
  considering formulas as games on which proofs induce strategies. We are
  specifically interested here in relating two such semantics of linear logic,
  of very different flavor, which both take in account concurrent features of
  the proofs: asynchronous games and concurrent games. Interestingly, we show
  that associating a concurrent strategy to an asynchronous strategy can be seen
  as a semantical counterpart of the focusing property of linear logic.
\end{abstract}

\vspace{-3ex}

A cut-free proof in sequent calculus, when read from bottom up, progressively
introduces the connectives of the formula that it proves, in the order specified
by the syntactic tree constituting the formula, following the conventions
induced by the logical rules. In this sense, a formula can be considered as a
playground that the proof will explore. The formula describes the rules that
this exploration should obey, it can thus be abstractly considered as a
\emph{game}, whose moves are its connectives, and a proof as a \emph{strategy}
to play on this game. If we follow the principle given by the Curry-Howard
correspondence, and see a proof as some sort of program, this way of considering
proof theory is particularly interesting because the strategies induced by
proofs describe very precisely the interactive behavior of the corresponding
program in front of its environment.

This point of view is at the heart of \emph{game semantics} and has proved to be
very successful in order to provide denotational semantics which is able to
describe precisely the dynamics of proofs and programs. In this interactive
perspective, two players are involved: the \emph{Proponent}, which represents
the proof, and the \emph{Opponent}, which represents its environment. A formula
induces a game which is to be played by the two players, consisting of a set of
moves together with the rules of the game, which are formalized by the polarity
of the moves (the player that should play a move) and the order in which the
moves should be played. The interaction between the two players is formalized by
a \emph{play}, which is a sequence of moves corresponding to the part of the
formula being explored during the cut-elimination of the proof with another
proof. A proof is thus described in this setting by a \emph{strategy} which
corresponds to the set of interactions that the proof is willing to have with
its environment.

This approach has been fruitful for modeling a wide variety of logics and
programming languages. By refining Joyal's category of Conway
games~\cite{joyal:conway} and Blass' games~\cite{blass:degrees-indeterminacy},
Abramsky and Jagadeesan were able to give the first fully complete game model of
the multiplicative fragment of linear logic extended with the MIX
rule~\cite{abramsky-jagadeesan:games-full-completeness}, which was later refined
into a fully abstract model of PCF (Programming Language of Computable
Functions)~\cite{abramsky-jagadeesan-malacaria:full-abstraction-pcf}. Here,
``fully complete'' and ``fully abstract'' essentially mean that the model is
very precise, in the sense that every strategy is \emph{definable} (\ie is the
interpretation of a proof or a program); more details can be found in Curien's
survey on the subject~\cite{curien:definability-fa}. Giving such a precise model
of this language, introduced by Plotkin~\cite{plotkin:lcf}, was considered as a
corner stone in computer science because it is a prototypical programming
language, consisting of the $\lambda$\nbd{}calculus extended with base data
types and a fixpoint operator. At exactly the same time, Hyland and Ong gave
another fully abstract model of PCF based on a variant of game semantics called
\emph{pointer games}~\cite{hyland-ong:full-abstraction-pcf}. In this model,
definable strategies are characterized by two conditions imposed on strategies
(well-bracketing and innocence). This setting was shown to be extremely
expressive: relaxing in various ways these conditions gave rise to fully
abstract models of a wide variety of programming languages with diverse features
such as references, control, etc.

Game semantics is thus helpful to understand how logic and typing regulate
computational processes. It also provides ways to analyze them (for example by
doing model checking~\cite{abramsky-ghica-murawski-ong:gs-verification}) or to
properly extend them with new features~\cite{curien:definability-fa}, and this
methodology should be helpful to understand better concurrent programs. Namely,
concurrency theory being relatively recent, there is no consensus about what a
good process calculus should be (there are dozens of variants of the
$\pi$\nbd{}calculus and only one $\lambda$\nbd{}calculus) and what a good typing
system for process calculus should be: we believe that the study of denotational
semantics of those languages is necessary in order to reveal their fundamental
structures, with a view to possibly extending the Curry-Howard correspondence to
programming languages with concurrent features. A few game models of concurrent
programming languages have been constructed and studied. In particular, Ghica
and Murawski have built a fully abstract model of Idealized Algol (an imperative
programming language with references) extended with parallel composition and
mutexes~\cite{ghica-murawski:angelic-semantics} and Laird a game semantics of a
typed asynchronous~$\pi$\nbd{}calculus~\cite{laird:pi-calc}.

In this paper, we take a more logical point of view and are specifically
interested in concurrent denotational models of linear logic. The idea that
multiplicative connectives express concurrent behaviors is present since the
beginnings of linear logic: it is namely very natural to see a proof of~$A\llpar
B$ or~$A\lltens B$ as a proof of~$A$ in ``parallel'' with a proof of~$B$, the
corresponding introduction rules being
\vspace{-1ex}
\[
  \inferrule{\vdash\Gamma,A,B}{\vdash\Gamma,A\llpar B}{\lrule{\llpar}}
  \qqtand
  \inferrule{\vdash\Gamma,A\\\vdash\Delta,B}{\vdash\Gamma,\Delta,A\lltens B}{\lrule{\lltens}}
\vspace{-1ex}
\]
with the additional restriction that the two proofs should be ``independent'' in
the case of the tensor, since the corresponding derivations in premise of the
rule are two disjoint subproofs. Linear logic is inherently even more parallel:
it has the \emph{focusing} property~\cite{andreoli:focusing} which implies that
every proof can be reorganized into one in which all the connectives of the same
polarity at the root of a formula are introduced at once (this is sometimes also
formulated using \emph{synthetic connectives}). This property, originally
discovered in order to ease proof-search has later on revealed to be fundamental
in semantics and type theory. Two game models of linear logic have been
developed in order to capture this intuition. The first, by Abramsky and
Melliès, called \emph{concurrent games}, models strategies as closure
operators~\cite{abramsky-mellies:concurrent-games} following the
domain-theoretic principle that computations add information to the current
state of the program (by playing moves). It can be considered as a big-step
semantics because concurrency is modeled by the ability that strategies have to
play multiple moves at once. The other one is the model of \emph{asynchronous
  games} introduced by Melliès~\cite{mellies:ag2}
where, in the spirit of ``true concurrency'', playing moves in parallel is
modeled by the possibility for strategies to play any interleaving of those
moves and these interleavings are considered to be equivalent. We recall here
these two models and explain here that concurrent games can be related to
asynchronous games using a semantical counterpart of focusing. A detailed
presentation of these models together with the proofs of many properties evoked
in this paper can be found in~\cite{mellies-mimram:ag5,mimram:phd}.

\vspace{-3ex}
\section{Asynchronous games}
\label{sec:ag}
\vspace{-2ex}
Recall that a \emph{graph}~$G=(V,E,s,t)$ consists of a set~$V$ of vertices (or
\emph{positions}), a set~$E$ of edges (or \emph{transitions}) and two
functions~$s,t:E\to V$ which to every transition associate a position which is
called respectively its \emph{source} and its \emph{target}. We
write~$m:x\transition{}y$ to indicate that~$m$ is a transition with~$x$ as
source and~$y$ as target. A \emph{path} is a sequence of consecutive transitions
and we write~$t:x\transitionpath{}y$ to indicate that~$t$ is a path whose source
is~$x$ and target is~$y$. The concatenation of two consecutive
paths~$s:x\transitionpath{}y$ and~$t:y\transitionpath{}z$ is denoted~$s\cdot
t$. An \emph{asynchronous graph}~$G=(G,\tile{})$ is a graph~$G$ together with a
\emph{tiling} relation~$\tile{}$, which relates paths of length two with the
same source and the same target. If~$m:x\transition{}y_1$,
$n:x\transition{}y_2$, $p:y_1\transition{}z$ and~$q:y_2\transition{}z$ are four
transitions, we write
\vspace{-2ex}
\begin{equation}
  \label{eq:tile}
  \vxym{
    &z&\\
    y_1\ar[ur]^p&\sim&y_2\ar[ul]_q\\
    &\ar[ul]^mx\ar[ur]_n&\\
  }
\vspace{-1ex}
\end{equation}
to diagrammatically indicate that~$m\cdot p\tile{}n\cdot q$. We write~$\sim$ for
the smallest congruence (\wrt concatenation) containing the tiling
relation. This relation is called \emph{homotopy} because it should be thought
as the possibility, when~$s$ and~$t$ are two homotopic paths, to deform
``continuously'' the path~$s$ into~$t$. From the concurrency point of view, a
homotopy between two paths indicates that these paths are the same up to
reordering of independent events, as in Mazurkiewicz traces. In the
diagram~\eqref{eq:tile}, the transition~$q$ is the \emph{residual} (in the sense
of rewriting theory) of the transition~$m$ after the transition~$n$, and
similarly $p$ is the residual of~$n$ after~$m$; the \emph{event} (also called
\emph{move}) associated to a transition is therefore its equivalence class under
the relation identifying a transition with its residuals. In the asynchronous
graphs we consider, we suppose that given a path~$m\cdot p$ there is at most one
path~$n\cdot q$ forming a tile~\eqref{eq:tile}. We moreover require that a
transition should have at most one residual after another transition.

We consider formulas of the multiplicative and additive fragment of linear logic
(MALL), which are generated by the grammar
\[
A\qgramdef A\llpar A\gramor A\lltens A\gramor A\llwith A\gramor A\llplus A\gramor X\gramor X^*
\vspace{-1ex}
\]
where~$X$ is a variable (for brevity, we don't consider units). The~$\llpar$
and~$\llwith$ (\resp $\lltens$ and~$\llplus$) connectives are sometimes called
\emph{negative} or \emph{asynchronous} (resp. \emph{positive} or
\emph{synchronous}). A \emph{position} is a term generated by the following
grammar
\vspace{-1ex}
\[
x\qgramdef \emptypos{}\gramor x\llpar x\gramor x\lltens x\gramor \llwithl
x\gramor \llwithr x\gramor \llplusl x\gramor \llplusr x
\vspace{-1ex}
\]
The de Morgan dual~$A^*$ of a formula is defined as usual, for example
$(A\otimes B)^*=A^*\llpar B^*$, and the dual of a position is defined
similarly. Given a formula~$A$, we write~$\fpos{A}$ for the set of valid
positions of the formula which are defined inductively by
$\emptypos{}\in\fpos{A}$ and if $x\in\fpos{A}$ and~$y\in\fpos{B}$ then \hbox{$x\llpar
y\in\fpos{A\llpar B}$}, $x\lltens y\in\fpos{A\lltens B}$, $\llwithl
x\in\fpos{A\llwith B}$, $\llwithr y\in\fpos{A\llwith B}$, $\llplusl
x\in\fpos{A\llplus B}$ and $\llplusr y\in\fpos{A\llplus B}$.

An \emph{asynchronous game} $G=(G,\initpos{},\lambda)$ is an asynchronous
graph~$G=(V,E,s,t)$ together with a distinguished \emph{initial position}
$\initpos{}\in V$ and a function \hbox{$\lambda:E\to\{O,P\}$} which to every
transition associates a \emph{polarity}: either $O$ for Opponent or $P$ for
Proponent. A transition is supposed to have the same polarity as its residuals,
polarity is therefore well-defined on moves. We also suppose that every
position~$x$ is \emph{reachable} from the initial position, \ie that there
exists a path~$\initpos{}\transitionpath{}x$. Given a game~$G$, we write~$G^*$
for the game~$G$ with polarities inverted. Given two games~$G$ and~$H$, we
define their asynchronous product~$G\|H$ as the game whose positions
are~$V_{G\|H}=V_G\times V_H$, whose transitions are~$E_{G\|H}=E_G\times
V_H+V_G\times E_H$ (by abuse of language we say that a transition is ``in~$G$''
when it is in the first component of the sum or ``in~$H$'' otherwise) with the
source of~$(m,x)\in E_G\times V_H$ being~$(s_G(m),x)$ and its target
being~$(t_G(m),x)$, and similarly for transitions in $V_G\times E_H$, two
transitions are related by a tile whenever they are all in~$G$ (\resp in $H$)
and the corresponding transitions in~$G$ (\resp in~$H$) are related by a tile or
when two of them are an instance of a transition in~$G$ and the two other are
instances of a transition in~$H$, the initial position is
$(\initpos{G},\initpos{H})$ and the polarities of transitions are those induced
by~$G$ and~$H$.

To every formula~$A$, we associate an asynchronous game~$G_A$ whose vertices are
the positions of~$A$ as follows. We suppose fixed the interpretation of the free
variables of~$A$. The game~$G_{A\llpar B}$ is obtained from the game~$G_A\|G_B$
by replacing every pair of positions~$(x,y)$ by~$x\llpar y$, and adding a
position~$\emptypos{}$ and an Opponent
transition~$\emptypos{}\transition{}\emptypos{}\llpar\emptypos{}$. The
game~$G_{A\llwith B}$ is obtained from the disjoint union \hbox{$G_A+G_B$} by
replacing every position~$x$ of~$G_A$ (\resp $G_B$) by~$\llwithl x$ (\resp
$\llwithr x$), and adding a position~$\emptypos{}$ and two Opponent transitions
$\emptypos{}\transition{}\llwithl\emptypos{}$ and
$\emptypos{}\transition{}\llwithr\emptypos{}$. The games associated to the other
formulas are deduced by de Morgan duality: $G_{A^*}=G_A^*$. This operation is
very similar to the natural embedding of event structures into asynchronous
transition systems~\cite{winskel-nielsen:models-concur}. For example, if we
interpret the variable~$X$ (\resp $Y$) as the game with two
positions~$\emptypos{}$ and~$x$ (\resp $\emptypos{}$ and~$y$) and one transition
between them, the interpretation of the formula~$(X\otimes X^*)\llwith Y$ is
depicted in~\eqref{eq:ex-intp}. We have made explicit the positions of the games
in order to underline the fact that they correspond to partial explorations of
formulas, but the naming of a position won't play any role in the definition of
asynchronous strategies.
\vspace{-2ex}
\begin{equation}
  \label{eq:ex-intp}
  \vxym{
    &\llwithl(x\lltens x^*)&\\
    \ar[ur]\llwithl(x\lltens\emptypos{})&\sim&\llwithl(\emptypos{}\lltens x^*)\ar[ul]\\
    &\llwithl(\emptypos{}\lltens\emptypos{})\ar[ul]\ar[ur]&&\llwithr y\\
    &\llwithl\emptypos{}\ar[u]&&\llwithr\emptypos{}\ar[u]\\
    &&\ar[ul]\emptypos{}\ar[ur]&\\
  }
\vspace{-1.5ex}
\end{equation}

A \emph{strategy}~$\sigma$ on a game~$G$ is a prefix-closed set of \emph{plays},
which are paths whose source is the initial position of the game. To every
proof~$\pi$ of a formula~$A$, we associate a strategy, defined inductively on
the structure of the proof. Intuitively, these plays are the explorations of
formulas allowed by the proof. For example, the strategies interpreting the
proofs
\vspace{-1ex}
\[
\inferrule
{
\inferrule{\pi}
{\vdash\Gamma,A,B}
}
{\vdash\Gamma,A\llpar B}{\lrule{\llpar}}
\qquad\qqtand\qquad
\inferrule
{
\inferrule{\pi}
{\vdash\Gamma,A}
}
{\vdash\Gamma,A\llplus B}{\lrule{\llplus_L}}
\vspace{-1ex}
\]
will contain plays which are either empty or start with a transition
$\emptypos{}\transition{}\emptypos{}\llpar\emptypos{}$ (\resp
$\emptypos{}\transition{}\llplusl\emptypos{}$) followed by a play in the
strategy interpreting~$\pi$. The other rules are interpreted in a similar
way. To be more precise, since the interpretation of a proof depends on the
interpretation of its free variables, the interpretation of a proof will be an
uniform family of strategies indexed by the interpretation of the free variables
in the formula (as in \eg\cite{abramsky-jagadeesan:games-full-completeness}) and
axioms proving~$\vdash A,A^*$ will be interpreted by copy-cat strategies on the
game interpreting~$A$. For the lack of space, we will omit details about
variables and axioms.

Properties characterizing definable strategies were studied in the case of
alternating strategies (where Opponent and Proponent moves should alternate
strictly in plays) in~\cite{mellies:ag4} and generalized to the non-alternating
setting that we are considering here in~\cite{mellies-mimram:ag5,mimram:phd}. We
recall here the basic properties of definable strategies. One of the interest of
these is that they allow one to reason about strategies in a purely local and
diagrammatic fashion. It can be shown that every definable strategy~$\sigma$ is
\vspace{-1ex}
\begin{itemize}
\item \emph{positional}: for every three paths
  \hbox{$s,t:\initpos{}\transitionpath{}x$} and~$u:x\transitionpath{}y$, if
  $s\cdot u\in\sigma$, $s\sim t$ and $t\in\sigma$ then $t\cdot u\in\sigma$. This
  property essentially means that a strategy is a subgraph of the game: a
  strategy~$\sigma$ induces a subgraph~$G_\sigma$ of the game which consists of
  all the positions and transitions contained in at least one play in~$\sigma$,
  conversely every play in this subgraph belongs to the strategy when the
  strategy is positional. In fact, this graph~$G_\sigma$ may be seen itself as
  an asynchronous graph by equipping it with the tiling relation induced by the
  underlying game.
\item \emph{deterministic}: if the graph~$G_\sigma$ of the strategy contains a
  transition~\hbox{$n:x\transition{}y_2$} and a Proponent
  transition~$m:x\transition{}y_1$ then it also contains the residual of~$m$
  along~$n$, this defining a tile of the form~\eqref{eq:tile}.
\item \emph{receptive}: if~$\sigma$ contains a
  play~$s:\initpos{}\transitionpath{}x$ and there exists an Opponent
  move~$m:x\transition{}y$ in the game then the play~$s\cdot
  m:\initpos{}\transitionpath{}y$ is also in~$\sigma$.
\item \emph{total}: if~$\sigma$ contains a play~$s:\initpos{}\transitionpath{}x$
  and there is no Opponent transition \hbox{$m:x\transition{}y$} in the game
  then either the position~$x$ is terminal (there is no transition with~$x$ as
  source in the game) or there exists a Proponent transition~$m:x\transition{}y$
  such that~$s\cdot m$ is also in~$\sigma$.
\end{itemize}



\vspace{-4ex}
\section{Focusing in linear logic}
\label{sec:focusing}
\vspace{-1ex}
In linear logic, a proof of the form depicted on the left-hand side of
\vspace{-1ex}
\[
\inferrule{
  \inferrule{
    \inferrule{
      \pi_1
    }{
      \vdash A,B,C
    }
  }{
    \vdash A\llpar B,C
  }{\lrule{\llpar}}
  \\
  \inferrule{
    \pi_2
  }{
    \vdash D
  }
}{
  \vdash A\llpar B,C\lltens D
}{\lrule{\lltens}}
\qquad\qquad\qquad\qquad
\inferrule{
  \inferrule{
    \inferrule{
      \pi_1
      }{
        \vdash A,B,C
      }
      \\
      \inferrule{
        \pi_2
      }{
        \vdash D
      }
  }{
    \vdash A,B,C\lltens D
  }{\lrule{\lltens}}
}{
  \vdash A\llpar B,C\lltens D
}{\lrule{\llpar}}
\vspace{-1ex}
\]
can always be reorganized into the proof depicted on the right-hand side. This
proof transformation can be seen as ``permuting'' the introduction of $\lltens$
after the introduction of~$\llpar$ (when looking at proofs bottom-up). From the
point of view of the strategies associated to the proofs, the game corresponding
to the proven sequent contains
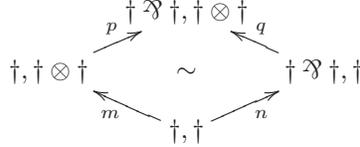
\vspace{-2ex}
\[
\vxym{
  &\emptypos{}\llpar\emptypos{},\emptypos{}\lltens\emptypos{}&\\
  \ar[ur]^p\emptypos{},\emptypos{}\lltens\emptypos{}&\sim&\emptypos{}\llpar\emptypos{},\emptypos{}\ar[ul]_q\\
  &\ar[ul]^m\emptypos{},\emptypos{}\ar[ur]_n&\\
}
\vspace{-1ex}
\]
and the transformation corresponds to replacing the path~$m\cdot p$ by the
path~$n\cdot q$ in the strategy associated to the proof. More generally, the
introduction rules of two negative connectives can always be permuted, as well
as the introduction of two positive connectives, and the introduction rule of a
positive connective can always be permuted after the introduction rule of a
negative one. Informally, a negative (\resp positive) can always be ``done
earlier'' (\resp ``postponed''). We write~$\pi\focleq\pi'$ when a proof~$\pi'$
can be obtained from a proof~$\pi$ by a series of such permutations of rules.

These permutations of rules are at the heart of Andreoli's
work~\cite{andreoli:focusing} which reveals that if a formula is provable then
it can be found using a \emph{focusing} proof search, which satisfies the
following discipline: if the sequent contains a negative formula then a negative
formula should be decomposed (\emph{negative phase}), otherwise a positive
formula should be chosen and decomposed repeatedly until a (necessarily unique)
formula is produced (\emph{positive phase}) -- this can be formalized using a
variant of the usual sequent rules for linear logic. From the point of view of
game semantics, this says informally that every strategy can be reorganized into
one playing alternatively a ``bunch'' of Opponent moves and a ``bunch'' of
Proponent moves.

All this suggests that proofs in sequent calculus are too sequential: they
contain inessential information about the ordering of rules, and we would like
to work with proofs modulo the congruence generated by the~$\focleq$
relation. Semantically, this can be expressed as follows. A strategy~$\sigma$ is
\emph{courteous} when for every tile of the form~\eqref{eq:tile} of the game,
such that the path~$m\cdot p$ is in (the graph~$G_\sigma$ of) the
strategy~$\sigma$, and either~$m$ is a Proponent transition or~$p$ is an
Opponent transition, the path~$n\cdot q$ is also in~$\sigma$. We
write~$\deseq\sigma$ for the smallest courteous strategy
containing~$\sigma$. Courteous strategies are less sequential than usual
strategies: suppose that~$\sigma$ is the strategy interpreting a proof~$\pi$ of
a formula~$A$, then a play~$s$ is in~$\deseq\sigma$ if and only if it is a play
in the strategy interpreting some proof~$\pi'$ such that~$\pi\focleq\pi'$.

Strategies which are positional, deterministic, receptive, total, courteous, are
closed under residuals of transitions and satisfy some other properties such as
the \emph{cube property} (enforcing a variant of the domain-theoretic stability
property) are called \emph{ingenuous} and are very well behaved: they form a
compact closed category, with games as objects and ingenuous strategies~$\sigma$
on~$A^*\| B$ as morphisms \hbox{$\sigma:A\to B$}, which is a denotational model
of MLL, which can be refined into a model of MALL by suitably quotienting
morphisms. Composition of strategies~$\sigma:A\to B$ and~\hbox{$\tau:B\to C$} is
defined as usual in game semantics by ``parallel composition and hiding'': the
plays in~$\tau\circ\sigma$ are obtained from \emph{interactions} of~$\sigma$
and~$\tau$, which are the plays on the game~$A\|B\|C$ whose projection
on~$A^*\|B$ (\resp $B^*\|C$) is in~$\sigma$ (\resp $\tau$) up to polarities of
moves,
by restricting them to~$A^*\|C$.
Associativity of the composition is not trivial and relies mainly on the
determinism property, which implies that if a play in~$\tau\circ\sigma$ comes
from two different interactions~$s$ and~$t$ then there it also comes from a
third interaction~$u$ which is greater than both \wrt the prefix modulo homotopy
order.



\vspace{-2.5ex}
\section{Concurrent games}
\vspace{-1.5ex}
We recall here briefly the model of concurrent
games~\cite{abramsky-mellies:concurrent-games}. A \emph{concurrent
  strategy}~$\varsigma$ on a complete lattice~$(D,\leq)$ is a continuous closure
operator on this lattice. Recall that a closure operator is a
function~$\varsigma:D\to D$ which is
\vspace{-1ex}
\begin{enumerate}
\item \emph{increasing}: $\forall x\in D, x\leq\varsigma(x)$
\item \emph{idempotent}: $\forall x\in D, \varsigma\circ\varsigma(x)=\varsigma(x)$
\item \emph{monotone}: $\forall x,y\in D, x\leq
  y\To\varsigma(x)\leq\varsigma(y)$
\end{enumerate}
\vspace{-1ex}
Such a function is \emph{continuous} when it preserves joins of directed
subsets. Informally, an order relation~$x\leq y$ means that the position~$y$
contains more information than~$x$. With this intuition in mind, the first
property expresses the fact that playing a strategy increases the amount of
information in the game, the second that a strategy gives all the information it
can given its knowledge in the current position (so that if it is asked to play
again it does not have anything to add), and the third that the more information
the strategy has from the current position the more information it has to
deliver when playing.

Every such concurrent strategy~$\varsigma$ induces a set of fixpoints defined as
the set \hbox{$\fix\varsigma=\setof{x\in D\tq \varsigma(x)=x}$}. This set is
(M)~closed under arbitrary meets and (J)~closed under joins of directed subsets
and conversely, every set~$X\subseteq D$ of positions which satisfies these two
properties (M) and (J) induces a concurrent strategy~$\clop{X}$ defined by
\hbox{$\clop{X}(x)=\bigwedge\setof{y\in X\tq x\leq y}$}, whose set of fixpoints
is precisely~$X$.

Suppose that~$G$ is a game. Without loss of generality, we can suppose that~$G$
is \emph{simply connected}, meaning that every position is reachable from the
initial position~$\initpos{}$ and two plays $s,t:\initpos{}\transitionpath{}x$
with the same target are homotopic. This game induces a partial order on its set
of positions, defined by $x\leq y$ iff there exists a
path~$x\transitionpath{}y$, which can be completed into a complete lattice~$D$
by formally adding a top element~$\top$. Now, consider a strategy~$\sigma$ on
the game~$G$. A position~$x$ of the graph~$G_\sigma$ induced by~$\sigma$ is
\emph{halting} when there is no Proponent move $m:x\transition{}y$ in~$\sigma$:
in such a position, the strategy is either done or is waiting for its Opponent
to play. It can be shown that the set~$\hpos\sigma$ of halting positions of an
ingenuous strategy~$\sigma$ satisfies the properties (M) and (J) and thus
induces a concurrent strategy~$\clop{(\hpos\sigma)}$. Conversely, if for every
positions~$x,y\in D$ we write~$x\leqP y$ when $y\neq\top$ and there exists a
path~$x\transitionpath{}y$ containing only Proponent moves, then every
concurrent strategy~$\varsigma$ induces a strategy~$\astrat\varsigma$ defined as
the set of plays in~$G$ whose intermediate positions~$x$ satisfy
$x\leqP\varsigma(x)$ -- and these can be shown to be ingenuous. This thus
establishes a precise relation between the two models:

\vspace{-1.5ex}
\begin{theorem}
  The two operations above describe a retraction of the ingenuous strategies on
  a game~$G$ into the concurrent strategies on the domain~$D$ induced by the
  game~$G$.
\end{theorem}

\vspace{-1.5ex}
\noindent Moreover, the concurrent strategies which correspond to ingenuous
strategies can be characterized directly. In this sense, concurrent strategies
are close to the intuition of focused proofs: given a position~$x$, they play at
once many Proponent moves in order to reach the position which is the image
of~$x$.

However, the correspondence described above is not functorial: it does not
preserve composition. This comes essentially from the fact that the category of
ingenuous strategies is compact closed, which means that it has the same
strategies on the games interpreting the formulas~$A\lltens B$ and~$A\llpar B$
(up to the polarity of the first move). For example, if~$X$ (\resp~$Y$) is
interpreted by the game with one Proponent
transition~$\emptypos{}\transition{}x$ (\resp $\emptypos{}\transition{}y$), the
interpretations of~$X^*\llpar Y^*$ and~$X\lltens Y$ are respectively
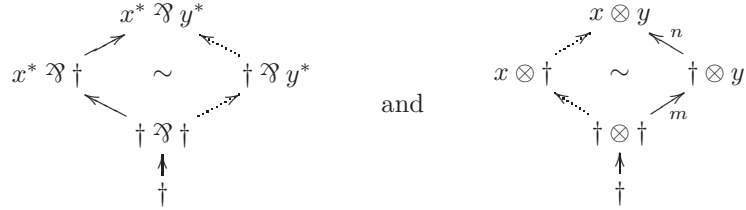
\vspace{-1ex}
\[
\vxym{
&x^*\llpar y^*&\\
\ar[ur]x^*\llpar\emptypos{}&\sim&\emptypos{}\llpar y^*\ar@{.>}[ul]\\
&\ar[ul]\emptypos{}\llpar\emptypos{}\ar@{.>}[ur]&\\
&\emptypos{}\ar[u]&
}
\qqtand
\vxym{
&x\lltens y&\\
\ar@{.>}[ur]x\lltens\emptypos{}&\sim&\emptypos{}\lltens y\ar[ul]_n\\
&\ar@{.>}[ul]\emptypos{}\lltens\emptypos{}\ar[ur]_m&\\
&\emptypos{}\ar[u]&
}
\vspace{-1ex}
\]
Now, consider the strategy~$\sigma:X^*\llpar Y^*$ which contains only the
prefixes of the bold path~$\emptypos{}\transitionpath{}(x^*\llpar y^*)$ and the
strategy~$\tau:X\lltens Y$ which contains only the prefixes of bold
path~$\emptypos{}\transitionpath{}(x\lltens y)$. The fixpoints of the
corresponding concurrent games are
respectively~$\hpos\sigma=\setof{\emptypos{},\ \emptypos{}\llpar\emptypos{},\
  x^*\llpar\emptypos{},\ x^*\llpar y^*}$ and \hbox{$\hpos\tau=\setof{x\lltens
    y}$}. From the point of view of asynchronous strategies, the only reachable
positions by both of the strategies in~$X\lltens Y$ are~$\emptypos{}$ and
$\emptypos{}\lltens\emptypos{}$. However, from the point of view of the
associated concurrent strategies, they admit the position~$x\lltens y$ as a
common position in~$X\lltens Y$. From this observation, it is easy to build two
strategies~$\sigma:A\to B$ and~$\tau:B\to C$ such that
\hbox{$\clop{(\hpos{(\tau\circ\sigma)})}\neq\clop{(\hpos\tau)}\circ\clop{(\hpos\sigma)}$}
(we refer the reader to~\cite{abramsky-mellies:concurrent-games} for the
definition of the composition of closure operators). In the example above, the
strategy~$\tau$ is the culprit: as mentioned in the introduction, the two
strategies on~$X$ and~$Y$ should be independent in a proof of~$X\lltens Y$,
whereas here the strategy~$\tau$ makes the move~$n$ depends on the
move~$m$. Formally, this dependence expresses the fact that the move~$m$ occurs
after the move~$n$ in every play of the
strategy~$\tau$. In~\cite{mellies-mimram:ag5}, we have introduced a
\emph{scheduling criterion} which dynamically enforces this independence between
the components of a tensor: a strategy satisfying this criterion is essentially
a strategy such that in a sub-strategy on a formula of the form~$A\lltens B$ no
move of~$A$ depends on a move of~$B$ and vice versa. Every definable strategy
satisfies this criterion and moreover,

\begin{theorem}
  Strategies satisfying the scheduling criterion form a subcategory of the
  category of ingenuous strategies and the
  operation~$\sigma\mapsto\clop{(\hpos\sigma)}$ extends into a functor from this
  category to the category of concurrent games.
\end{theorem}

This property enables us to recover more precisely the focusing property
directly at the level of strategies as follows. Suppose that~$\sigma$ is an
ingenuous strategy interpreting a proof~$\pi$ of a
sequent~$\vdash\Gamma$. Suppose moreover that \hbox{$s:x\transitionpath{}y$} is
a maximal play in~$\sigma$. By receptivity and courtesy of the strategy, this
play is homotopic in the graph~$G_\sigma$ to the concatenation of a
path~$s_1:x\transitionpath{}x_1$ containing only Opponent moves, where~$x_1$ is
a position such that there exists no Opponent
transition~$m:x_1\transition{}x_1'$, and a path~$s_2:x_1\transitionpath{}y$.
Similarly, by totality of the strategy, the path~$s_2$ is homotopic to the
concatenation of a path~$s_2':x_1\transitionpath{}y_1$ containing only Proponent
moves, where~$y_1$ is a position which is either terminal or such that there
exists an Opponent transition~$m:y_1\transition{}y_1'$, and a
path~$s_2'':y_1\transitionpath{}y$. The path~$s_2'$ consists in the partial
exploration of positive formulas, one of them being explored until a negative
subformula is reached. By courtesy of the strategy, Proponent moves permute in a
strategy and we can suppose that~$s_2'$ consists only in such a maximal
exploration of one of the formulas available at the position~$x$. If at some
point a branch of a tensor formula is explored, then by the scheduling criterion
it must be able to also explore the other branch of the formula. By repeating
this construction on the play~$s_2''$, every play of~$\sigma$ can be transformed
into one which alternatively explores all the negative formulas and explores one
positive formula until negative formulas are reached. By formalizing further
this reasoning, one can therefore show that

\begin{theorem}
  \label{thm:foc}
  In every asynchronous strategy interpreting a proof in MALL is included a
  strategy interpreting a focusing proof of the same sequent.
\end{theorem}

A motivation for introducing concurrent games was to solve the well-known
\emph{Blass problem} which reveals that the ``obvious'' game model for the
multiplicative fragment of linear logic has a non-associative
composition. Abramsky explains in~\cite{abramsky:sequentiality-concurrency} that
there are two ways to solve this: either syntactically by considering a focused
proof system or semantically by using concurrent games. Thanks to asynchronous
games, we understand here the link between the two points of view: every proof
in linear logic can be reorganized into a focused one, which is semantically
understood as a strategy playing multiple moves of the same polarity at once,
and is thus naturally described by a concurrent strategy. In this sense,
concurrent strategies can be seen as a semantical generalization of focusing,
where the negative connectives are not necessarily all introduced at first
during proof-search. It should be noted that some concurrent strategies are less
sequential than focused ones, for example the strategies interpreting the
multi-focused proofs~\cite{chaudhuri-miller-saurin:multi-focusing}, where the
focus can be done on multiple positives formulas in the positive phase of the
focused proof-search. Those multi-focused proofs were shown to provide canonical
representatives of focused proofs (the interpretation of the canonical
multi-focused proof can be recovered by generalizing Theorem~\ref{thm:foc}). We
are currently investigating a generalization of this result by finding canonical
representatives for concurrent strategies.

The author is much indebted to Paul-André Melliès and Martin Hyland.



\vspace{-3ex}

\bibliographystyle{plain}
\bibliography{these}

\end{document}